\documentclass[12pt]{article}
\input epsf

\newcommand{\bea}{\begin{eqnarray}}
\newcommand{\eea}{\end{eqnarray}}
\newcommand{\be}{\begin{equation}}
\newcommand{\ee}{\end{equation}}
\newcommand{\pa}{\partial}
\newcommand{\bpa}{\bar{\partial}}
\newcommand{\nn}{\nonumber \\}
\newcommand{\bz}{\bar{z}}

\makeatletter

\@addtoreset{equation}{section}
\makeatother
\def\href#1#2{#2}

\begin{document}

\begin{titlepage}

\vspace*{30mm}

\begin{center}
{\LARGE From Free Fields to $AdS$ -- Thermal Case}\\
\vspace*{18mm}
{\scshape  FURUUCHI \ Kazuyuki}\\
\vspace*{2mm}
{\sl Harish-Chandra Research Institute}\\
{\sl Chhatnag Rd., Jhusi, Allahabad 211 019, India}\\
\vspace*{1mm}
{\tt furuuchi@mri.ernet.in}
\vspace*{3mm}
\begin{abstract}
We analyze the
reorganization of free field theory correlators
to closed string amplitudes
investigated in 
\cite{Gopakumar:2003ns,Gopakumar:2004qb,%
Gopakumar:2005fx,Gopakumar:2004ys}
in the case of Euclidean thermal field theory
and study how the dual bulk geometry 
is encoded on them.
The expectation value of Polyakov loop,
which is an order parameter for
confinement-deconfinement transition,
is directly reflected on the 
dual bulk geometry.
The dual geometry
of confined phase is found to be 
$AdS$ space
periodically identified in Euclidean time direction.
The gluing of Schwinger parameters,
which is a key step for the
reorganization of field theory correlators,
works in the same way as in the
non-thermal case.
In deconfined phase
the gluing is made possible only by taking
the dual geometry correctly.
The dual geometry for
deconfined phase
does not have a non-contractible
circle in the Euclidean time direction.
\end{abstract}
\end{center}

\end{titlepage}       

\tableofcontents

\section{Introduction}

The large $N$ gauge theory-closed string
duality conjecture has been providing
us a lot of deep
insights into both gauge theories
and gravity.
Yet despite of 
the recent developments
in some specific
examples 
\cite{Gopakumar:1998ki,Gaiotto:2003yb,%
Ooguri:2002gx,Berkovits:2003pq,%
McGreevy:2003kb,McGreevy:2003ep,%
Teschner:2005rd},
we still don't know how to
directly translate the descriptions in
one side to the other.

Maldacena's conjecture
\cite{Maldacena:1997re,Gubser:1998bc,Witten:1998qj}
suggests that weakly coupled 
large $N$
conformal field theories
are dual to closed string theories on highly curved 
$AdS$ space.
String theory
on $AdS$ space is not yet
developed sufficiently
to test the Maldacena's conjecture 
in a complete precision,
but the tractability of field theory in
weak coupling 
makes us hope that it will be possible
to construct a string theory directly from
a field theory in this limit.%
\footnote{%
There appeared a lot of literature
on this weak/free field theory-string theory
correspondence recently.
See
\cite{Sundborg:1999ue,Haggi-Mani:2000ru,Sundborg:2000wp,%
Polyakov:2001af,%
Karch:2002vn,Clark:2003wk,%
Bonelli:2003zu,Bonelli:2004ve,%
Tseytlin:2002gz,%
Dhar:2003fi,Dolan:2003uh,%
Sezgin:2002rt,%
Beisert:2003te,Bianchi:2004xi,%
Okuyama:2004bd,Alday:2005nd}
for other approaches in this direction.
There is also a lot of literature
on the connection
between weakly coupled ${\cal N}=4$
Yang-Mills theory and integral spin chain
since the work of \cite{Minahan:2002ve}.
For a recent approach to
the reorganization of field theory correlators
to string worldsheets in light-cone gauge,
see \cite{Bardakci:2001cn,Thorn:2002fj}.}

In a series of papers
\cite{Gopakumar:2003ns,Gopakumar:2004qb,%
Gopakumar:2005fx,Gopakumar:2004ys},
taking free field theory as
a starting point
Gopakumar 
presented a refined 
argument 
of 't Hooft's reorganization \cite{'tHooft:1973jz}
of large $N$
field theory correlators to closed string amplitudes,
and made more precise study 
of the gauge theory-closed string 
duality possible.
The key step was gluing
of propagators in each edge of a
given Feynman diagram
to make up a so called skeleton graph.
Each edge is assigned one 
effective Schwinger parameter
after the gluing.
Then a mathematical result
\cite{Harer,Penner,Kontsevich:1992ti}
tells us that the 
space of Schwinger parameters
gives cell decomposition of
${\cal M}_{g,n} \times R_+^n$,
where ${\cal M}_{g,n}$
is a moduli space of genus $g$ Riemann surface
with $n$ punctures.
The appearance of ${\cal M}_{g,n}$
is a strong support
that the large $N$ field theory correlators
do organize themselves to closed
string amplitudes.
The resulting closed string 
is propagating in
$AdS$ geometry,
as expected from the conformal 
symmetry \cite{Maldacena:1997re}.

It is interesting to test the generality
of the method to see how
the information of string theory,
for example difference of backgrounds,
is encoded on field theory correlators. 
In particular, recently thermodynamics
of weakly coupled Yang-Mills theories
have been studied extensively
\cite{Sundborg:1999ue,Polyakov:2001af,%
Aharony:2003sx,Aharony:2005bq,%
Furuuchi:2003sy,%
Hadizadeh:2004bf,%
Schnitzer:2004qt,%
Spradlin:2004pp,Spradlin:2004sx,%
Gomez-Reino:2005bq}
and it is interesting to ask what
are the dual geometries in these cases.\footnote{%
We understand that B. Rai has also raised 
the question of understanding the thermal geometry
from the field theory.}
Although there are qualitative agreements
between phase transitions in Yang-Mills theories
and expected phase transitions in corresponding
bulk geometries,
since weakly coupled gauge theories
are dual to highly curved space-times
with curvature scales being the order of string scale
one cannot regorously analyze
the geometry without string theory.
For this reason
the reorganization of field theory correlators
into closed string amplitudes is a promising
approach to study these 
highly stringy geometries.

In this article, we analyze the
reorganization of free field theory correlators
in Euclidean thermal field theory.
Two typical phases of 
gauge field theories at finite temperature
are confined phase and
deconfined phase.
The order parameter of 
the phase transition
is Polyakov loop.
We find that 
the expectation value of Polyakov loop
is crucial for reconstructing the
bulk geometry.
In section \ref{Confined}
we study confined phase.
The dual geometry
of confined phase is found to be an
$AdS$ space periodically identified in
Euclidean time direction
(thermal $AdS$), as expected.
However, this is not just a simple
consequence of the periodic identification
in field theory side alone.
The fact that the expectation value of the 
Polyakov loop is zero in confined phase
is crucial for
reconstructing the bulk $AdS$ geometry.
The gluing of Schwinger parameters,
which was a key step for
reorganizing field theory correlators
to closed string amplitudes
\cite{Gopakumar:2004qb,Gopakumar:2005fx},
works in the same way as in the
non-thermal case.
In section \ref{Deconfined}
we turn to deconfined phase.
It turns out that the gluing of Schwinger parameters
does not work straightforwardly
in deconfined phase.
We analyze the reason 
and identified it with a behavior of
"string bits"
\cite{Giles:1977mp,Klebanov:1988ba}, 
which is reminiscent of that in
the Hagedorn transition
\cite{Sathiapalan:1986db,O'Brien:1987pn,%
Kogan:1987jd,Atick:1988si}. 
We will argue that we can 
nevertheless glue
the field theory correlators if
we take the dual geometry correctly
and examine the meaning of the gluing 
in more general context. 
The dual geometry of deconfined phase
is not easily found in general,
but in section \ref{Example}
we present a simple example
where we can explicitely 
find the dual bulk geometry.
This is a two dimensional
CFT on $S^1\times R$,
where $S^1$ is the thermal circle.
Our general arguments on
deconfined phase and its dual geometry
in section \ref{Deconfined}
are concretely realized in this example.

\section{Confined Phase vs. Bulk Geometry}\label{Confined}

In this section we will explain
how in confined phase
field theory correlators see
the dual bulk geometry.%
\footnote{%
Discussions in this section 
apply to interacting field theories also,
regardless of the title of this article.
Confinement is usually regarded as a strong coupling
phenomenon and it is somewhat counterintuitive
to study it in the free field limit 
which is a
main focus of this article.
However, it has been shown
that if one takes
(\ref{zeroP})
below
(or 
$\langle |{\cal P}|^2 \rangle = 0$
for gauge theories on
a compact spacial manifold)
as a
criterion for confinement it does
occur in some weakly coupled gauge theories
on a compact manifold at large $N$.
A good explanation on this account
is given in \cite{Aharony:2003sx},
see also the summary \cite{Aharony:2004ir}.}
The result is rather simple:
If they see some dual bulk geometry
at zero temperature,
at finite temperature
they just see the same 
geometry 
with periodic identification 
in Euclidean time direction.
However, this is not a simple consequence
of the periodic identification
in field theory side alone.
The expectation value of Polyakov loop
plays a crucial role.
The expectation value of Polyakov loop
is an order parameter
of confinement-deconfinement
transition, and 
in $AdS$/CFT correspondence
it has a dual description
in terms of string worldsheet in the bulk 
whose end is on the loop
\cite{Rey:1998ik,Maldacena:1998im,%
Witten:1998zw,Rey:1998bq,Brandhuber:1998bs}.
It is interesting to observe
how Polyakov loop
directly reflects the bulk 
geometry in our approach.

For concreteness,
let us study 
massless scalar field $\Phi$
in adjoint representation of
gauge group $SU(N)$
on $S^1\times R^{d-1}$,
where $S^1$ is the thermal circle
parameterized by $\tau$ with period $\beta$.
We will work in the 
gauge where $A_0$ is constant and diagonal.
A criterion for confinement
is that the Polyakov loop expectation value 
vanishes:
\bea
\label{zeroP}
\langle{\cal P}\rangle =0, 
\quad
{\cal P} \equiv
\frac{1}{N}\mbox{Tr}P \, \exp i 
\int_0^{\beta} d\tau A_0
\eea
where $P$ denotes the path ordering.
\footnote{In the case where the spacial
manifold is compact rather than
$R^{d-1}$, 
$\langle |{\cal P}|^2\rangle$ is a more suitable
order parameter \cite{Aharony:2003sx}.}
This is realized
by the following configuration:
\bea
\label{sym}
A_0 =
\frac{2\pi}{\beta N}
\left(
\mbox{diag}
(1,\cdots ,N)
- \frac{N+1}{2}
\right).
\eea
In the 't Hooft limit 
$N\rightarrow \infty$, 
$g_{YM}\rightarrow 0$
with $\lambda = g_{YM}^2 N$ fixed,
the action is generically 
of order $N^2$
whereas there are $N$
diagonal components of $A_0$,
so fluctuations of them
around saddle points are surpressed.
Whether the configuration
(\ref{sym}) is realized 
as the most dominant saddle point
depends on 
the theory 
(other matter contents etc.).
However, once (\ref{sym}) is realized
the following argument can be applied
regardless of the details of the theory.
So we assume (\ref{sym})
and study its consequence.
Let us study 
perturbative expansion
around (\ref{sym}).%
\footnote{%
We thank S. Minwalla for stressing us that
the difference of the configuration
of $A_0$ should be reflected when we probe
the bulk geometry 
by field theory correlators.}
The quadratic part of the action 
in the presence of $A_0$ 
zero-mode configuration (\ref{sym}) is
\bea
&&\frac{1}{g_{YM}^2}
\int_0^\beta d\tau\int d^{d-1}x 
D_\mu \Phi_{ab} D^\mu \Phi_{ba} \nonumber \\
&=& 
\frac{1}{g_{YM}^2 \beta}
\sum_{n=-\infty}^\infty \int d^{d-1}p 
\left(
 \left(\frac{n + (A_{0\, aa}-A_{0\, bb})}{\beta/2\pi}\right)^2 + p^2
\right)
\Phi_{ab}(\frac{2\pi n}{\beta},p)
\Phi_{ba}(-\frac{2\pi n}{\beta},-p)
\nn
&=&
\frac{1}{g_{YM}^2 \beta}
\sum_{n=-\infty}^\infty \int d^{d-1}p
\left(
 \left(\frac{n + \frac{a-b}{N}}{\beta/2\pi}\right)^2 + p^2
\right)
\Phi_{ab}(\frac{2\pi n}{\beta},p)\Phi_{ba}(-\frac{2\pi n}{\beta},-p).
\eea
Thus the propagator is
\bea
\label{propagator}
\langle
\Phi_{ab}(n,p) \Phi_{cd}(-n,-p)
\rangle_{S^1 \times R^{d-1}}
=
\delta_{ad}\delta_{bc}
\frac{g_{YM}^2}{\left(\frac{2\pi}{\beta}(n + \frac{a-b}{N})\right)^2+p^2}.
\eea
We have included the effect of the 
$A_0$ zero-mode configuration (\ref{sym})
in the propagator.\footnote{%
The techniques below are 
reminiscent of those used in the reduced models 
\cite{Eguchi:1982nm,Parisi:1982gp,Gross:1982at,%
Bhanot:1982sh,Das:1982ux}.}

\begin{figure}
\begin{center}
 \leavevmode
 \epsfxsize=50mm
 \epsfbox{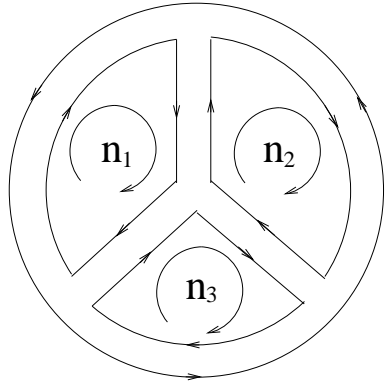}\\
\end{center}
\caption{Assigning loop momenta to matrix index loops
in 't Hooft's double line representation
of gauge theory planar Feynman diagram.
In planar diagram the loop momenta are one less than the index loops, so
there remains one extra index loop 
which we chose to be the outer index loop here.}
\label{loopi}
\end{figure}

Since we are considering 
planar limit
$N \rightarrow \infty$,
we can replace matrix index sums
by integrals:
\bea
 \label{indexint}
\sum_{a_i=1}^N f(a_i)
\rightarrow 
\frac{\beta N}{2\pi}
\int_0^{\frac{2\pi}{\beta}} dp_{0 i}\, f(p_{0 i}),
\eea
for some function $f(a_i)$.
Notice the factor of $N$ in front 
of the integral.
This means that the modification
of the propagator (\ref{propagator})
by the
zero-mode of the gauge field $A_0$
(\ref{sym})
does not change the argument of 't Hooft:
Each matrix index loop, or face,
contributes with a factor of $N$.
Now suppose we calculate correlation functions
in this theory.
In planar diagrams, we can parameterize
the temporal loop momenta 
by matrix index line notation 
(see e.g. \cite{Gross:1982at,Minwalla:1999px}).
For a given planar Feynman diagram
with $\ell$ momentum loops,
we have $\ell+1$ matrix index loop.
We assign the loop momentum $n_i$
$(i =1, \cdots, \ell)$ to every index loop
but one, say $(\ell+1)$-th index loop
(Fig.\ref{loopi}).
Correlation functions
are calculated by 
connecting the
fields and vertices
with the propagators (\ref{propagator}),
and $a_i$'s only appear
in the combination $a_i-a_j$.
This doesn't depend on a constant shift
to all $a_i$
with some integer $q$,
so one summation of internal indices, 
which we conventionally
choose to
be $a_{\ell+1}$,
gives just a factor $N$.
Also 
the index $p_{0i}$ and 
the loop momentum $n_i$ 
can always be combined
as
$p_{0i}+ \frac{2\pi n_i}{\beta}$.
The origin of this combination is
gauge covariance,
the covariant derivative in the Euclidean
time direction.
Notice that both momentum- and index- flows
are associated with a direction (indicated by an arrow)
in the 't Hooft's double line
representation of Feynman diagrams.
Thus we can combine all 
the other index loop sums, which were replaced
by integral in (\ref{indexint}), with the
temporal momentum sums
$n_i$
(\ref{indexint}):
\bea
\sum_{n_i=-\infty}^\infty
\int_0^{\frac{2\pi}{\beta}} dp_{0i}
f (p_{0i}+ \frac{2\pi n_i}{\beta}  )
=
\int_{-\infty}^{\infty} dp_{0i} f(p_{0i})
\eea
for some function $f(p_{0i})$.
Therefore internal loop momentum integrals
in calculation of correlation functions
on $S^1\times R^{d-1}$ with the
$A_0$ zero-mode configuration 
(\ref{sym})
become the same as 
the ones without the $S^1$ compactification:
Suppose we have a $M$-point function
of gauge invariant operators on $R\times R^{d-1}$
as a function of incoming momenta $k$
(we surpress the spatial momenta
in the following expressions):%
\footnote{All the correlation functions
studied in this article will be connected diagrams
and we will not explicitely mention that hereafter.
For non-connected diagrams
our arguments straightforwardly
apply to each connected components.}
\bea
\langle {\cal O}_1(k_{0 1})
\cdots {\cal O}_M (k_{0 M}) \rangle_{R\times R^{d-1}}
=
G_M (k_{0 1},\cdots k_{0 M}) .
\eea 
Then on $S^1 \times R^{d-1}$
we will obtain the same function $G_M$ 
with incoming momenta $\frac{2\pi m}{\beta}$:
\bea
\langle {\cal O}_1(\frac{2\pi m_1}{\beta})
\cdots {\cal O}_M(\frac{2\pi m_M}{\beta}) 
\rangle_{S^1\times R^{d-1}}
=
G_M (\frac{2\pi m_1}{\beta},\cdots ,
\frac{2\pi m_M}{\beta}) .
\eea
The only difference is that
the incoming momenta are discrete.
Let the Fourier transform of $G_M(k)$ be $G_M(\tau)$:
\bea
\frac{1}{\sqrt{2\pi}}
\int_{-\infty}^\infty dk_0 G_M(k_0) e^{ik_0\tau}
=
G_M(\tau)
\eea
where we schematically picked up 
one incoming momentum, but
the calculation is the same for all the incoming momenta.
Then by the Poisson resummation formula,
\bea
\frac{\sqrt{2\pi}}{\beta}
\sum_{m=-\infty}^{\infty}
G_M(\frac{2\pi m}{\beta}) e^{i \frac{2\pi m}{\beta} \tau}
=
\sum_{n=-\infty}^\infty
G_M(\tau + \beta n) .
\eea
This means if one reads off some 
geometry 
of the bulk from 
a field theory on 
$R \times R^{d-1}$,
on $S^1 \times R^{d-1}$ 
with zero Polyakov loop expectation value
one just finds the same 
geometry
except that it has a
periodic identification in $\tau$ direction.

Since 
correlation functions 
on $S^1 \times R^{d-1}$ 
in momentum space
have the same form as in the $R\times R^{d-1}$
case, the gluing procedure of
\cite{Gopakumar:2004qb} works
exactly in the same way in confined phase.

\begin{figure}
\begin{center}
 \leavevmode
 \epsfxsize=60mm
 \epsfbox{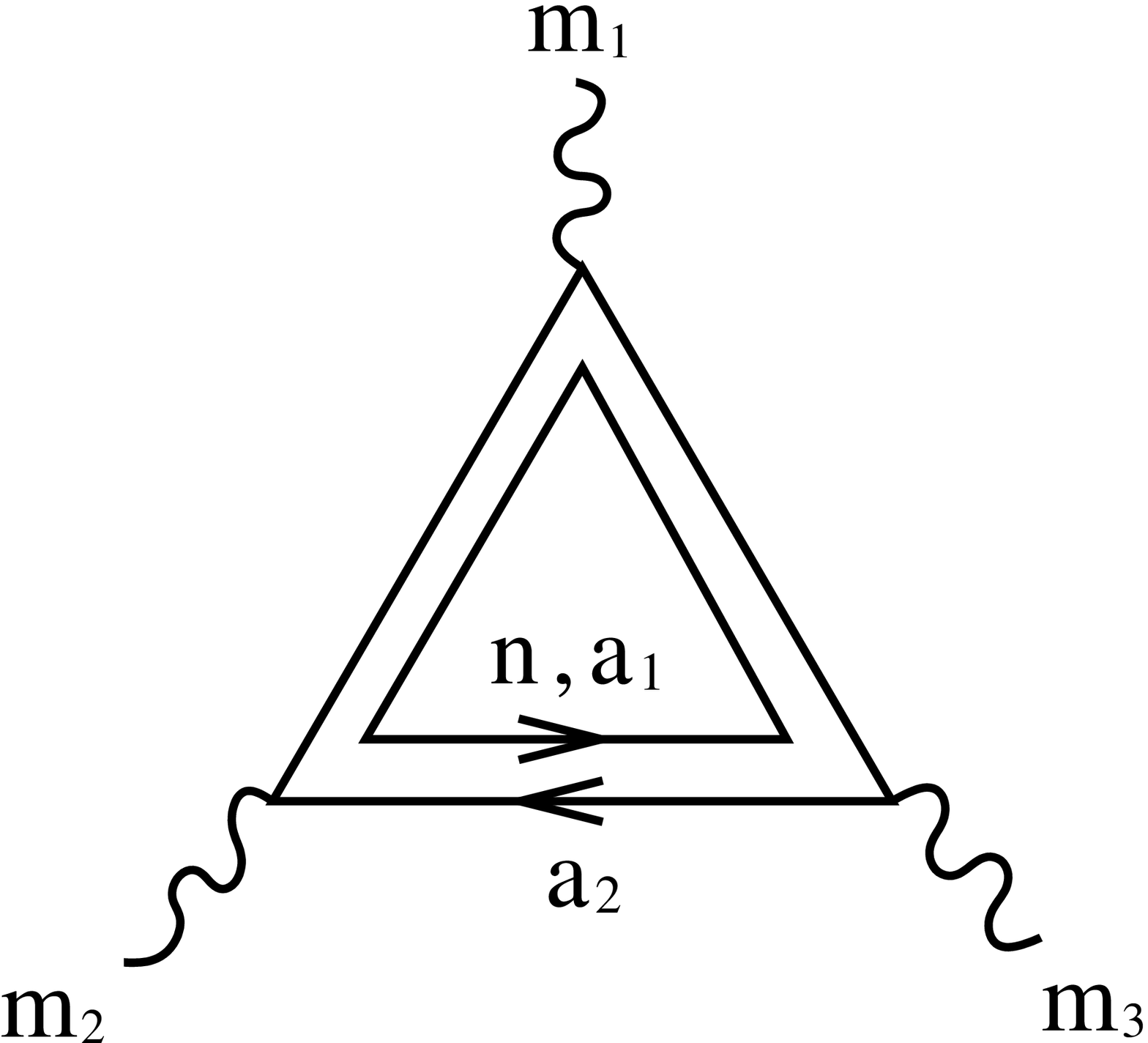}\\
\end{center}
\caption{}
\label{triangle}
\end{figure}

\begin{figure}
\begin{center}
 \leavevmode
 \epsfxsize120mm
 \epsfbox{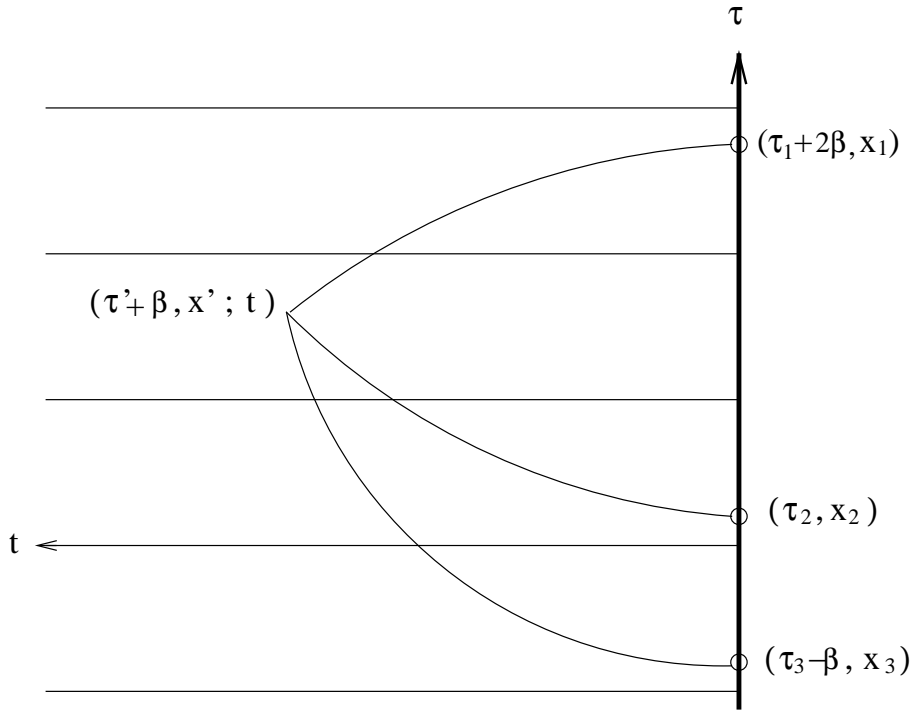}\\
\end{center}
\caption{A diagram contributing to (\ref{3ptpAdS}). 
The field theory correlator can be expressed 
in terms of the
bulk to boundary propagators in $AdS_{d+1}$
periodically identified in $\tau$ direction:
The bulk to boundary propagators (solid curves)
connect images of operators at
point $(\tau_1,x_1;0)$, $(\tau_2,x_2;0)$, $(\tau_3,x_3;0)$
and the bulk point $(\tau',x';t)$.
}
\label{periodic}
\end{figure}

As an example,
let us calculate the following simple
three point function 
in free field theory (Fig.\ref{triangle}):
\bea
\langle 
\mbox{Tr} \Phi^2(\frac{2\pi m_1}{\beta},k_1)
\mbox{Tr} \Phi^2(\frac{2\pi m_2}{\beta},k_2)
\mbox{Tr} \Phi^2(\frac{2\pi m_3}{\beta},k_3)
\rangle_{S^1\times R^{d-1}}
\eea
under the $A_0$ zero-mode configuration (\ref{sym}).%
\footnote{Actually confined phase is not
thermodynamically
favored in free field theory on
$S^1\times R^{d-1}$.
However, once the phase 
of a system of interest
is known to
be in confined phase
the following calculation 
does not depend
on the details of the spatial manifold
or interactions.
Therefore please regard this example
as an exhibition of the calculational essence
applicable for field theories
which do have confined phase.}
Up to the total momentum conservation delta function
$\delta_{m_1+m_2+m_3,0}\delta(k_1+k_2+k_3)$
this is 
\bea
&&
\sum_{a_1=1}^N\sum_{a_2=1}^N
\frac{1}{\beta}
\sum_{n=-\infty}^\infty
\int d^{d-1}p \nonumber \\
&&\frac{g_{YM}^6}{%
\left(\left(\frac{2\pi n + \frac{a_1-a_2}{N}}{\beta}\right)^2\!\!+p^2 \right)\!\!%
\left(\left(\frac{2\pi (n-m_2) + \frac{a_1-a_2}{N}}{\beta}\right)^2\!\!+(p-k_2)^2 \right)\!\!%
\left(\left(\frac{2\pi (n+m_3) + \frac{a_1-a_2}{N}}{\beta}\right)^2\!\!+(p+k_3)^2\right)%
}.\nonumber \\
\eea
Then 
we turn loop index 
$a_1$ summation
into integral $\int_0^{\frac{2\pi}{\beta}}dp_{0}$
and combine the $p_{0}$ integral with 
the temporal momentum 
$n$ summation.
The result is
\bea
\label{3ptr}
N^2\cdot \lambda^3 N^{-3}
\int dp_{0} d^{d-1}p
\frac{1}{%
(p_{0}^2+p^2)%
\left(\left(p_{0}-\frac{2\pi m_2}{\beta}\right)^2+(p-k_2)^2\right)%
\left(\left(p_{0}+\frac{2\pi m_3}{\beta}\right)^2+(p+k_3)^2\right)}. \nonumber\\
\eea
The conversion of the summation
over $n$ to the integral
gave a factor of $N$, and
redundant summation over $a_2$
gave another factor of $N$,
together with contributions
from propagators
resulting a factor of $N^2\cdot N^{-3}$
which is appropriate for 
sphere with three punctures.
(\ref{3ptr}) is the same form as the 
one in $R \times R^{d-1}$ case, 
except that the temporal momenta
take discrete values.
The three point function on $R \times R^{d-1}$
in position space
can be written as \cite{Gopakumar:2004qb}
\bea
&&\langle
\mbox{Tr} \Phi^2(\tau_1,x_1)
\mbox{Tr} \Phi^2(\tau_2,x_2)
\mbox{Tr} \Phi^2(\tau_3,x_3)
\rangle_{R \times R^{d-1}} \nonumber \\
&=&
\lambda^3 N^{-1}
\int_0^\infty \frac{dt}{t^{\frac{d}{2}+1}}
\int_{-\infty}^\infty d\tau' 
\int_{-\infty}^\infty d^{d-1}x' 
\prod_{s=1}^3 K_{\Delta}(\tau_s,x_s;\tau',x';t),
\eea
where
\bea
K_{\Delta}(\tau,x;\tau',x';t)
=
\frac{t^{\frac{\Delta}{2}}}{[t+(\tau-\tau')^2+(x-x')^2]^\Delta}
\eea
is the usual position space bulk to boundary propagator
in $AdS_{d+1}$ for a scalar field corresponding
to an operator of dimension $\Delta$,
$\Delta = d-2$ in the case at hand.
Then on $S^1 \times R^{d-1}$
we get
\bea
&&\langle
\mbox{Tr} \Phi^2(\tau_1,x_1)
\mbox{Tr} \Phi^2(\tau_2,x_2)
\mbox{Tr} \Phi^2(\tau_3,x_3)
\rangle_{S^1 \times R^{d-1}} \nonumber \\
&=&
\lambda^3 N^{-1}
\int_0^\infty \frac{dt}{t^{\frac{d}{2}+1}}
\int_{-\infty}^\infty d\tau' 
\int_{-\infty}^\infty d^{d-1}x' 
\prod_{s=1}^3 
\sum_{n_s=-\infty}^\infty
K_{\Delta}(\tau_s + \beta n_s,x_s;\tau',x';t) \nonumber \\
&=&
\lambda^3 N^{-1}
\int_0^\infty \frac{dt}{t^{\frac{d}{2}+1}}
\int_0^\beta d\tau' 
\int_{-\infty}^\infty d^{d-1}x' 
\prod_{s=1}^3 
\sum_{n=-\infty}^\infty
\sum_{n_s=-\infty}^\infty
K_{\Delta}(\tau_s + \beta n_s,x_s;\tau'+\beta n,x';t).
\nonumber \\
\label{3ptpAdS}
\eea
In (\ref{3ptpAdS}) we sum over images of 
the bulk to boundary
propagators in $AdS_{d-1}$
periodically identified in $\tau$ direction 
(Fig.\ref{periodic}), as we have
explained in more general settings.
The confined phase is not 
actually thermodynamically
favored in free field theory
on $S^1 \times R^d$.
However, the above calculation did not depend
on the details of the spatial directions.
Therefore it can be straightforwardly
applied to the case of other spatial manifolds,
for example to free Yang-Mills theory
on $S^3$ extensively studied in 
\cite{Aharony:2003sx,Aharony:2005bq}.
There the confined phase is realized
at low temperature.
In this case one just needs 
to replace $p^2$ 
above to Laplacian for conformally coupled scalars
on $S^3$,
and momentum integration to
sum over spherical harmonics on $S^3$
with an appropriate measure factor.

Before closing this section,
we emphasize again that
a periodic identification in
field theory side
does not by itself lead to
the periodic identification 
in the dual bulk geometry.
The $A_0$ zero-mode configuration (\ref{sym}),
which tells that the system under consideration
is in confined phase, 
is crucial
for finding the periodically identified 
bulk geometry. 

\section{Deconfined Phase vs. Bulk Geometry}\label{Deconfined}

\begin{figure}
\begin{center}
 \leavevmode
 \epsfxsize=110mm
 \epsfbox{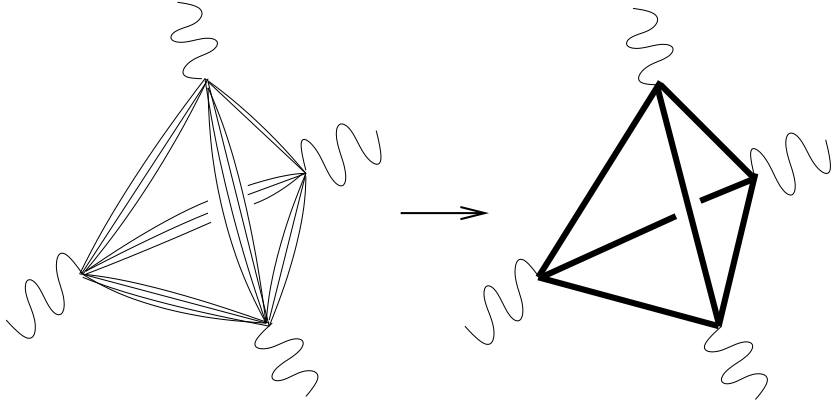}\\
\end{center}
\caption{Gluing propagators
so that each edge is parameterized by one Schwinger parameter.}
\label{gluing}
\end{figure}

Now let us turn to deconfined phase.
We will study the
case where 
the zero-mode of $A_0$ is zero
so that
the expectation value of the Polyakov
loop is one.
\footnote{This may be the most 
typical configuration, but 
this is not the most general case.}
Let us take free massless 
adjoint scalar field $\Phi$ on
$S^1\times R^{d-1}$ as an example.
For later purpose, let us work in position space.
The propagator on $S^1\times R^{d-1}$
in position space representation
can be obtained by summing over
images on its covering space $R\times R^{d-1}$:
\bea
\label{images}
\langle 
\Phi_{ab}(\tau,x)\Phi_{cd}(0) 
\rangle_{S^1\times R^{d-1}}
=
\sum_{n=-\infty}^\infty
\langle 
\Phi_{ab}(\tau+\beta n,x)\Phi_{cd}(0) 
\rangle_{R\times R^{d-1}}.
\eea
The propagator on $R\times R^{d-1}$ is given by
\bea
\label{propbit}
\langle 
\Phi_{ab}(\tau,x)\Phi_{cd}(0) 
\rangle_{R\times R^{d-1}}
=
\delta_{ad}\delta_{bc}
\frac{1}{(\tau^2+x^2)^\frac{d-2}{2}} .
\eea
Now let us recall the argument of 
\cite{Gopakumar:2004qb}
for reorganizing field theory correlators
into closed string amplitudes.
The crucial step was
the gluing of propagators in each edge
of a given Feynman diagram
to make up a so-called skeleton graph (Fig.\ref{gluing}).

The gluing in position space is seen
as follows.\footnote{%
We thank R. Gopakumar
for stressing the usefulness
of viewing the gluing in position space.}
Suppose the $r$-th edge has $m_r$ propagators.
Then, it has a contribution proportional to
\bea
\label{position}
\left(
\frac{1}{(\tau^2+x^2)^\frac{d-2}{2}}
\right)^{m_r}.
\eea
One can exponentiate each propagator
by a Schwinger parameter to obtain
\bea
\label{bitSch}
\left(
\frac{1}{(\tau^2+x^2)^\frac{d-2}{2}}
\right)^{m_r}
=
\prod_{\mu_r=1}^{m_r}
\frac{1}{\Gamma (\frac{d-2}{2})}
\int_0^\infty d\sigma_{\mu_r} 
\sigma_{\mu_r}^{\frac{d-2}{2}-1}
e^{-\sigma_{\mu_r}(\tau^2+x^2)}.
\eea
If we insert the identity 
\bea
1 = 
\int_0^\infty d\sigma_r
\delta(\sigma-\sum_{\mu_r=1}^{m_r}\sigma_{\mu_r})
\eea
to (\ref{bitSch}) and 
then change the variables to
$\alpha_{\mu_r}=\frac{\sigma_{\mu_r}}{\sigma_r}$,
we obtain
\bea
&&
\prod_{\mu_r=1}^{m_r}
\frac{1}{\Gamma (\frac{d-2}{2})}
\int_0^\infty d\sigma_r
\sigma_{r}^{\frac{d-2}{2}m_r}
e^{-\sigma_{r}(\tau^2+x^2)}
\int_0^1 d\alpha_{\mu_r} 
\delta(\sigma (1 - \sum_{\mu_r=1}^{m_r} \alpha_{\mu_r}))
\nonumber\\
&=&
\left(
\frac{1}{\Gamma (\frac{d-2}{2})}
\right)^{m_r}
\int_0^\infty d\sigma_r
\sigma_{r}^{\frac{d-2}{2}m_r-1}
e^{-\sigma_{r}(\tau^2+x^2)}
\cdot
\prod_{\mu_r=1}^{m_r}
\int_0^1 d\alpha_{\mu_r} 
\delta(1 - \sum_{\mu_r=1}^{m_r}\alpha_{\mu_r}).
\eea
The alpha integrals factor out
to give an overall constant
$
\frac{\left(\Gamma(\frac{d-2}{2})\right)^{m_r}}%
{\Gamma(\frac{d-2}{2}m_r)}
$
and one effective Schwinger parameter
$\sigma_r$ remains.
Thus we have "glued" 
Schwinger parameters 
into an effective
Schwinger parameter $\sigma_r$. 
By Fourier transforming it into momentum space
one gets the expression given in 
\cite{Gopakumar:2004qb} and
$\sigma_r$ is related to the momentum space
effective Schwinger parameter $\tau_r$ 
in \cite{Gopakumar:2004qb}
as 
$\sigma_r=\frac{1}{\tau_r}$.
On the other hand, one 
could have
exponentiated (\ref{position}) at once
using just one Schwinger parameter:
\bea
\label{Gamma}
\left(
\frac{1}{(\tau^2+x^2)^\frac{d-2}{2}}
\right)^{m_r}
=
\frac{1}{\Gamma (\frac{d-2}{2}m_r)}
\int_0^\infty d\sigma_r \sigma_r^{\frac{d-2}{2} m_r-1}
e^{-\sigma_r(\tau^2+x^2)}.
\eea
This seems to suggest that
in position space 
the multiplication of 
propagators 
itself may be
identified with the "gluing".

Once propagators in each edge
are glued and each edge 
in the resulting skeleton graph
is assigned a
Schwinger parameter, 
the space of 
the Schwinger parameters
gives a cell decomposition of
${\cal M}_{g,n}\times R_+^n$
as argued in \cite{Gopakumar:2004qb},
where ${\cal M}_{g,n}$ is a moduli space of
$n$-punctured genus $g$ Riemann surface.
The appearance of the moduli space
${\cal M}_{g,n}$ 
is a strong indication 
that large $N$ field theory amplitudes do
reorganize themselves into
closed strings. See 
\cite{Gopakumar:2003ns,Gopakumar:2004qb,%
Gopakumar:2005fx,Gopakumar:2004ys}
for more details.

How gluing works in the case
of $S^1 \times R^{d-1}$?
One may try to follow
the arguments of \cite{Gopakumar:2004qb}
which were done in momentum representation
(see also \cite{Itzykson:1980rh}).
However, the difference between 
$R \times R^{d-1}$
and 
$S^1 \times R^{d-1}$ manifests itself
in several stages
and one cannot straightforwardly follow the 
steps in the $R \times R^{d-1}$ case.
We let the interested readers
to try to follow the arguments of 
\cite{Gopakumar:2004qb} and
\cite{Itzykson:1980rh}
in the case of $S^1 \times R^{d-1}$ and
see where the differences appear
(however see the remarks below).
Here, instead, we work in position space and
try to understand the reason why
on $S^1 \times R^{d-1}$
gluing procedure
is not straightforward.
First, we would like to interpret
the propagator (\ref{propbit})
as a propagation of a 
"string bit" $\Phi$ in $AdS$ space
which makes up a closed string.
This is plausible 
since up to the matrix indices
the functional dependence
of the propagator is fixed by
conformal symmetry,
and the function of the form in
(\ref{propbit});
\bea
\frac{1}{(\tau^2+x^2)^\frac{d-2}{2}}
\eea
can always be understood
in terms of bulk to boundary propagator in $AdS_{d+1}$.
Then, if we use
the righthand side
of (\ref{images}) to calculate
correlators, it
may be interpreted as
summing over contributions of
each bit propagating in 
periodically identified $AdS$ space
going around the $S^1$ direction for
a different number of times.
In that case, the resulting
closed string worldsheet will be
wildly torn apart.
This is somewhat reminiscent of the interpretation
of the Hagedorn transition in string theory
\cite{Sathiapalan:1986db,O'Brien:1987pn,%
Kogan:1987jd,Atick:1988si}.
This will make the closed string interpretation in
{\em $AdS$ space} inappropriate in deconfined phase.
We identify this as
the reason why on $S^1\times R^3$
the gluing procedure is not straightforward.

\begin{figure}
\begin{center}
 \leavevmode
 \epsfxsize=80mm
 \epsfbox{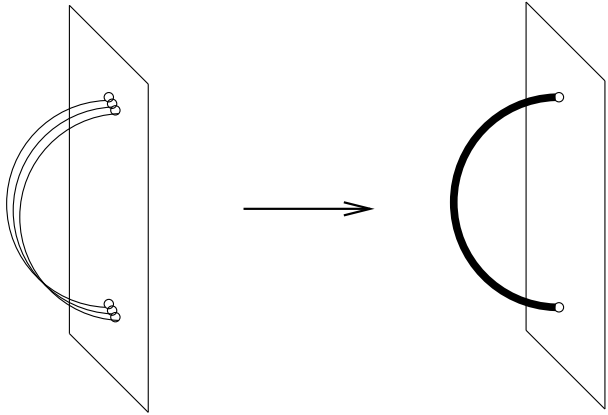}\\
\end{center}
\caption{The gluing of a two point function in 
the "new geometry" in
geodesic approximation.
The sum of the contributions
from the propagation of each string bit
can be expressed as contribution from 
a single geodesic with the effective mass 
being the sum of the mass of the string bits.}
\label{glugeo}
\end{figure}

However, now we would like to argue that
one can nevertheless "glue" the propagators 
and give a closed string interpretation,
but not in the $AdS$ space 
but in a {\em different} 
bulk geometry.
For that we interpret 
the {\em result} of the summation in (\ref{images})
as coming from
a bit propagator 
in this {\em new} bulk geometry.
This interpretation with a new bulk geometry
is also reminiscent of the speculation about
the phase after the Hagedorn transition
\cite{Barbon:2001di,Barbon:2002nw,Barbon:2004dd}.
Since there is no summation over images any more,
there should not be non-contaractable circle
in the new geometry.
This conclusion is in good accordance
with the criteria of confinement/deconfinement
from string worldsheet consideration
in the bulk \cite{Witten:1998zw}.
Then,
the gluing procedure in this case
will be 
multiplication
of position space propagators in each edge,
as was the case in $R \times R^{d-1}$.
Geodesic approximation illustraits
this interpretation:
For large $J$ it gives
\bea
\label{geodesicapp}
&&
\langle
\mbox{Tr} \Phi^J(p) \mbox{Tr} \Phi^J(q)
\rangle_{S^1\times R^{d-1}} 
\sim \langle \Phi(p) \Phi(q) \rangle_{S^1\times R^{d-1}}^J 
\nonumber \\
&\sim&
\left. e^{- M  D_{reg}(p,q)}
\right|_{{new\ geometry}}
= 
\left. 
\left(e^{- m D_{reg}(p,q)} 
\right)^J \right|_{{new\ geometry}}
+ {\cal O}(1/J) 
\eea
where
$M = \frac{1}{R}\sqrt{\Delta (\Delta-d)}$,
(with $\Delta = J(d-2)/2$) is a mass of the
bulk particle corresponding
to the operator 
$
\mbox{Tr} \Phi^J
$
according to the $AdS$/CFT dictionary
\cite{Gubser:1998bc,Witten:1998qj}.
$m = (d-2)/2R$
is interpreted as a mass of a string bit.
$D_{reg}(p,q)$ is a regularized
geodesic distance in the new geometry
between the points
$p$ and $q$ on the boundary.
The formula (\ref{geodesicapp}) 
tells us that
the gluing of string bit geodesics corresponds to
taking a single geodesic
with the effective mass given by
a sum of all the masses of the string bits,
which results in multiplying  
field theory propagators.
See Fig.\ref{glugeo}.

The reason why the gluing
didn't work straightforwardly in
momentum space
was as follows.
If one naively tries to follow the
argument of \cite{Gopakumar:2004qb}
in momentum space,
it means he/she is 
using the Fourier transform of (\ref{propbit})
as a propagator instead
of the lefthand side of (\ref{images}).
Then the gluing does not work,
at least straightforwardly,
as we have argued above.

Although we don't know a good 
parameterization
of Schwinger parameters which is
convenient for comparing with
a conjectural closed string theory amplitudes
on this new background geometry,%
\footnote{In the case of
field theory on $R^d$  
whose dual is supposed to be a
closed string theory
on $AdS_{d+1}$,
the correspondence between
Schwinger parameters and the 
moduli space of string worldsheet has
largely developed in the recent investigation
\cite{Gopakumar:2005fx}.}
one can certainly assign
one parameter to each edge, which we
also call "Schwinger parameter", 
once propagators in each edge are glued
to make a skeleton graph.

We will examine the above picture
by studying a simple example
in the next section.

\section{CFT on $S^1 \times R$ and the Dual Bulk Geometries}\label{Example}

\subsection{Free Scalar CFT on $S^1 \times R$}

In this section we will study
free massless scalar field theory on
$S^1 \times R$.
But before going into the $S^1 \times R$ case,
it is useful to recall the $R \times R$ case,
the covering space of $S^1 \times R$.
In particular, we will recall
how the geometry of $AdS$ space
is encoded in conformal
field theory correlators.

Let $\tau$ be the Euclidean time coordinate
and $x$
be the spatial coordinate on $R$.
We define
$z = \tau + i x$.
The correlation functions are given 
by\footnote{%
In two dimension the scalar field
$\Phi$ is not a conformal operator
so we will consider $\pa \Phi$ 
($\bpa \Phi$) instead.}
\bea
\langle
\pa \Phi_{ab}(z) \pa \Phi_{cd}(0)
\rangle_{R \times R}
=
\delta_{ad}\delta_{bc}
\frac{1}{z^2},
\label{RRz}
\eea
\bea
\langle
\bpa \Phi_{ab}(\bz) \bpa \Phi_{cd}(0)
\rangle_{R \times R}
=
\delta_{ad}\delta_{bc}
\frac{1}{\bz^2}.
\label{RRbz}
\eea
From the above, we can calculate 
two point functions like\footnote{%
For gauge invariance we may better use
covariant derivatives $D\Phi$ ($\bar{D}\Phi$).
However, we can choose the gauge $\pa_\tau A_0=0$,
$A_x = 0$,
and here we are considering the phase 
where the zero-mode of $A_0$ is zero.}
\bea
\label{ftresult}
\langle
\mbox{Tr} (\pa \Phi)^J (\bpa \Phi)^J (z,\bz)
\mbox{Tr} (\pa \Phi)^J  (\bpa \Phi)^J (0)
\rangle_{R \times R}
\sim
\left(
\frac{1}{z^2}
\right)^J
\left(
\frac{1}{\bz^2}
\right)^J .
\eea
This coinsides with
the geodesic approximation in $AdS_3$ space
\bea
 \label{1b}
ds^2
=
r^2 \left(\frac{d\tau^2 + dx^2}{\beta^2}\right)
+ R^2 \frac{dr^2}{r^2}
\eea
in large $J$ limit (see Appendix \ref{A}):
The geodesic approximation 
in this metric gives
\bea
\label{appgeo}
\langle
\mbox{Tr} (\pa \Phi)^J (\bpa \Phi)^J (z,\bz)
\mbox{Tr} (\pa \Phi)^J  (\bpa \Phi)^J (0)
\rangle_{R \times R}
\sim
\left(
\frac{1}{z^2}
\right)^{\sqrt{J(J-1)}}
\left(
\frac{1}{\bz^2}
\right)^{\sqrt{J(J-1)}} + {\cal O}(1/J) .
\eea
which coinsides with (\ref{ftresult})
up to ${\cal O}(1/J)$ terms.\footnote{%
One may use boundary to bulk propagators
instead of the geodesic
approximation 
to refine the match between field theory side and
bulk side in this example.
Since we are 
interested in the general picture 
we can extract 
rather than a particular nature of 
this simple geometry, 
we content ourselves 
with the geodesic approximation.}

Now let us study the 
free massless scalar field theory on
$S^1 \times R$, 
where the Euclidean time $\tau$
is compactified on $S^1$ with period $\beta$.
By "free",
we mean
the dimensionless parameter
$\beta^2 \lambda \rightarrow 0$,
where $\lambda = g_{YM}^2N$ is the
't Hooft coupling.%
This system is in deconfined phase
and the expectation value of the
zero-mode of $A_0$ is zero.
Thus the propagators on $S^1 \times R$
can be obtained by summing over
images of its universal covering $R \times R$:
\bea
\langle
\pa \Phi_{ab}(z) \pa \Phi_{cd}(0)
\rangle_{S^1 \times R}
&=&
\sum_{n=-\infty}^{\infty}
\langle
\pa \Phi_{ab}(z+ \beta n) \pa \Phi_{cd}(0)
\rangle_{R \times R} \\
&=&
\delta_{ad}\delta_{bc}
\sum_{n=-\infty}^{\infty}
\frac{1}{(z+ \beta n)^2} \label{g1z}\\
&=&
\delta_{ad}\delta_{bc}
\frac{\pi^2}{\beta^2 \sin^2 \frac{\pi z}{\beta}}
\label{g2z},
\eea
\bea
\langle
\bpa \Phi_{ab}(\bz) \bpa \Phi_{cd} (0)
\rangle_{S^1 \times R}
&=&
\sum_{n=-\infty}^{\infty}
\langle
\bpa \Phi_{ab}(\bz+ \beta n) \bpa \Phi_{cd} (0)
\rangle_{R \times R} \\
&=&
\delta_{ad}\delta_{bc}
\sum_{n=-\infty}^{\infty}
\frac{1}{(\bz+ \beta n)^2} \label{g1bz}\\
&=&
\delta_{ad}\delta_{bc}
\frac{\pi^2}{\beta^2 \sin^2 \frac{\pi \bz}{\beta}}
\label{g2bz} .
\eea
Using the above, typically we find 
two point functions like
\bea
\langle
\mbox{Tr} (\pa \Phi)^J (\bpa \Phi)^J (z,\bz)
\mbox{Tr} (\pa \Phi)^J  (\bpa \Phi)^J (0)
\rangle
\sim
\left(
\frac{1}{\cosh (\frac{2\pi x}{\beta}) 
- \cos  \frac{2\pi \tau}{\beta}}
\right)^{2J}.
\label{cft2}
\eea
In large $J$ limit this coinsides with the
result obtained from geodesic approximation
(see Appendix \ref{A}) in 
Euclidean $AdS_3$ metric in the static coordinates
(with its Euclidian time 
in $AdS$ sense being the $x$ direction)
\bea
\label{2pp}
ds^2
= 
r^2 \frac{d\tau^2}{\beta^2}
+
\left({r^2} + {R^2}\right)\frac{dx^2}{\beta^2}
+
\frac{dr^2}{\frac{r^2}{R^2}+1}.
\eea
It is important to notice that this geometry
is {\em different} from the geometry
(\ref{1b}) periodically identified in $\tau$
direction.
Although both are locally
$AdS$, the choice
of the Euclidean time direction
according to 
which the energy is defined is different
in both cases.
Also, the geometry (\ref{1b}) 
with the periodic identification
has a non-contractible
circle in $\tau$ direction
whereas (\ref{2pp}) does not.
(\ref{2pp}) is an example of what we called
"new geometry" in the
previous section.
In this simple example the new geometry
is again locally $AdS_3$, but it will be
different for different cases,
like free fields in
higher dimension etc.
In the next subsection
we will argue that this 
new geometry is thermodynamically
favored at finite temperature
in the saddle point
approximation of Euclidean path integral 
gravity, in accordance with
the discussions in the previous section.

\subsection{Zero Temperature 
Phase Transition in the Bulk}%

The boundary geometry $S^1\times R$
may admit two bulk 
saddle points, depending on the gravitational
action $I$ which is a functional of the
bulk metric.
We will
comment on the action $I$ shortly.
One saddle point will be
the Euclidean $AdS_3$ 
in the Poincare coordinates (\ref{1b})
with periodic identification in $\tau$ 
(thermal $AdS$):
\bea
\label{1}
ds^2
=
r^2 \left(\frac{d\tau^2 + dx^2}{\beta^2}\right)
+ R^2 \frac{dr^2}{r^2} .
\eea
The other is (\ref{2pp}), the Euclidean $AdS_3$ in
the static coordinates 
(with its Euclidian time in $AdS$ sense
being the $x$ direction):
\bea
\label{2p}
ds^2
= 
r^2 \frac{d\tau^2}{\beta^2}
+
\left({r^2} + {R^2}\right)\frac{dx^2}{\beta^2}
+
\frac{dr^2}{\frac{r^2}{R^2}+1}.
\eea
We can make a change of coordinate
$r^2 \rightarrow r^2-R^2$ in (\ref{2p})
to obtain the following form
\bea
\label{2}
ds^2=
(r^2-R^2)\frac{d\tau^2}{\beta^2}
+
r^2\frac{dx^2}{\beta^2}
+
\frac{dr^2}{\frac{r^2}{R^2}-1}.
\eea
The metric (\ref{2}) is a special case
of the one studied in \cite{Witten:1998zw},
which simply reduces to $AdS_3$ 
in the case of three bulk dimension.
We will call (\ref{1}) geometry I and
(\ref{2p}) or (\ref{2}) geometry II.
Note that the geometry II
would have a conical singularity
unless the period of $\tau$ is $\beta$.
The geometry I 
has a non-contractible circle
along $\tau$ direction (see Fig.\ref{nc}).
There is no surface with a disk topology in the bulk
that ends on a boundary thermal circle,
and from the dictionary of $AdS$/CFT correspondence
this corresponding to the
zero Polyakov loop expectation value
in the CFT side \cite{Witten:1998zw}.
The geometry II 
covers whole $AdS_3$ 
and there's no
non-contractible circle (see Fig.\ref{cc}), 
corresponding to the
non-zero Polyakov loop expectation value
in the CFT side.

In order to discuss
thermodynamics
in the Euclidean path integral formulation
of gravity,
we need to know the action $I$.
We expect this action to be 
ultimately
derived from a
conjectural closed string theory
dual to the free field theory.
The closed string loop correction
is surpressed by $\frac{1}{N}$
in the planar limit,
but the string $\alpha'$ correction
will be large since the
dual geometry has a curvature scale
around the order of the string scale.
Although we may expect $AdS_3$
to be an exact string background,
we do not know how the
string correction to the action would be.
However,
for a constant curvature space (locally $AdS_3$)
the classical gravitational action 
is proportional to the volume
if it is generally covariant, i.e.
made out of generally covariant
combinations of metric and 
curvature tensor and its covariant derivatives.
Schematically,
\bea
\qquad  \qquad \qquad \qquad \quad
I \propto \int dx^3 \sqrt{g}
\quad (\mbox{for constant curvature spaces}).
\label{action}
\eea
We assume the coefficient
for the proportionality is positive.
We also assume that 
the geometry I and II are the only
two minima of this action.
Actually, the above volume (\ref{action})
is infinite so
we need to regularize it.
In order for that we introduce a
cut off $r_{reg}$ in the radial coodrinate $r$
and take the difference of the two volumes
$V_{II}(r_{reg})$ and $V_{I}(r_{reg})$
corresponding to two geometries
II and I 
respectively
\cite{Hawking:1982dh,Witten:1998zw}:
\bea
V_{II}(r_{reg}) - V_{I}(r_{reg})
=
\frac{R}{\beta^2}
\int^{\beta}_0 d\tau
\int^{r_{reg}}_R dr
\int^L_0 dx \, \, r
-
\frac{R}{\beta^2}
\int^{\beta'}_0 d\tau
\int^{r_{reg}}_0 dr
\int^L_0 dx \, \, r .
\eea
We reqire the physical circumference
of the time direction to be equal at
$r=r_{reg}$
\bea
\beta \sqrt{r_{reg}^2-R^2}
=
\beta' r_{reg} .
\eea
Then the difference of the volumes per unit length
in $x$ direction is given by
\bea
\label{vol}
\lim_{r_{reg}\rightarrow \infty}
\frac{V_{II}(r_{reg}) - V_{I}(r_{reg})}{L}
=
- \frac{R^3}{4\beta}.
\eea
The above means we
use the 
geometry I as a reference point
to measure free energy.
This may be natural
because this is the
minimum free energy solution
at zero-temperature.
(\ref{vol}) means the action of the geometry II 
has negative free energy,
so in the saddle point approximation
as soon as we put the system in finite temperature 
the geometry II
is preferred.
This is in good accordance with
the thermodynamics of field theory side and 
the discussions in section {\ref{Deconfined}}.%
\footnote{%
Precisely speaking,
the reason it appears as a 
zero-temperature transition is that
we took the high temperature
limit $\beta^2 \lambda \rightarrow 0$.
The structure of the phase transition is
actually 
hidden at $\beta^2 \lambda \sim 1$
where our free field description is
not valid.
As long as we are interested in the high energy
phase $\beta^2 \lambda << 1$
we can use the free field description.}

The gluing, which was crucial for the closed string
picture, is possible only in the 
"new" geometry II
which corresponds to deconfined phase. 
Notice that 
in our approach one can see that
the instability of the geometry I
is a direct consequence of
the instability of the symmetric
configuration of the zero-mode
of the temporal gauge field $A_0$ (\ref{sym}):
Once the symmetric $A_0$
configuration ceases to be
stable,
field theory correlators
start seeing
a different geometry.

The energy density $E$ per unit length
of the geometry II 
is given by
\bea
E= \frac{\pa}{\pa \beta} I
\propto \frac{R^3}{4\beta^2}
= \frac{R^3}{4} T^2
\eea
and entropy density $S$ is
\bea
S = \beta E - I
\propto \frac{R^3}{2\beta}
= \frac{R^3}{2} T  .
\eea
Thus up to the coefficients
the geometry II reproduces the
results expected for
free field theory in two dimension.
So the assumptions on the action
(\ref{action}) qualitatively
reproduce the field theory results.

If we rescale the coordinate 
$r \rightarrow \frac{\beta}{R} r$,
the metric (\ref{1}) becomes
\bea
\label{1r}
ds^2 = \frac{r^2}{R^2}(d\tau^2+dx^2)
+ R^2 \frac{dr^2}{r^2}
\eea
whereas the metric (\ref{2}) becomes
\bea
\label{2r}
ds^2
= \left( \frac{r^2}{R^2} - \frac{R^2}{\beta^2}\right) d\tau^2
+ \frac{r^2}{R^2} dx^2
+ \frac{dr^2}{\frac{r^2}{R^2}-\frac{R^2}{\beta^2}}  .
\eea
In this form it is clear 
that in the low temperature
limit ($\beta \rightarrow \infty$)
the metric (\ref{2r}) reduces to the metric (\ref{1r}).

\subsection{Deconfinement, Hagedorn Transition, String Bits
and Gluing}

\begin{figure}
\begin{center}
 \leavevmode
 \epsfxsize=60mm
 \epsfbox{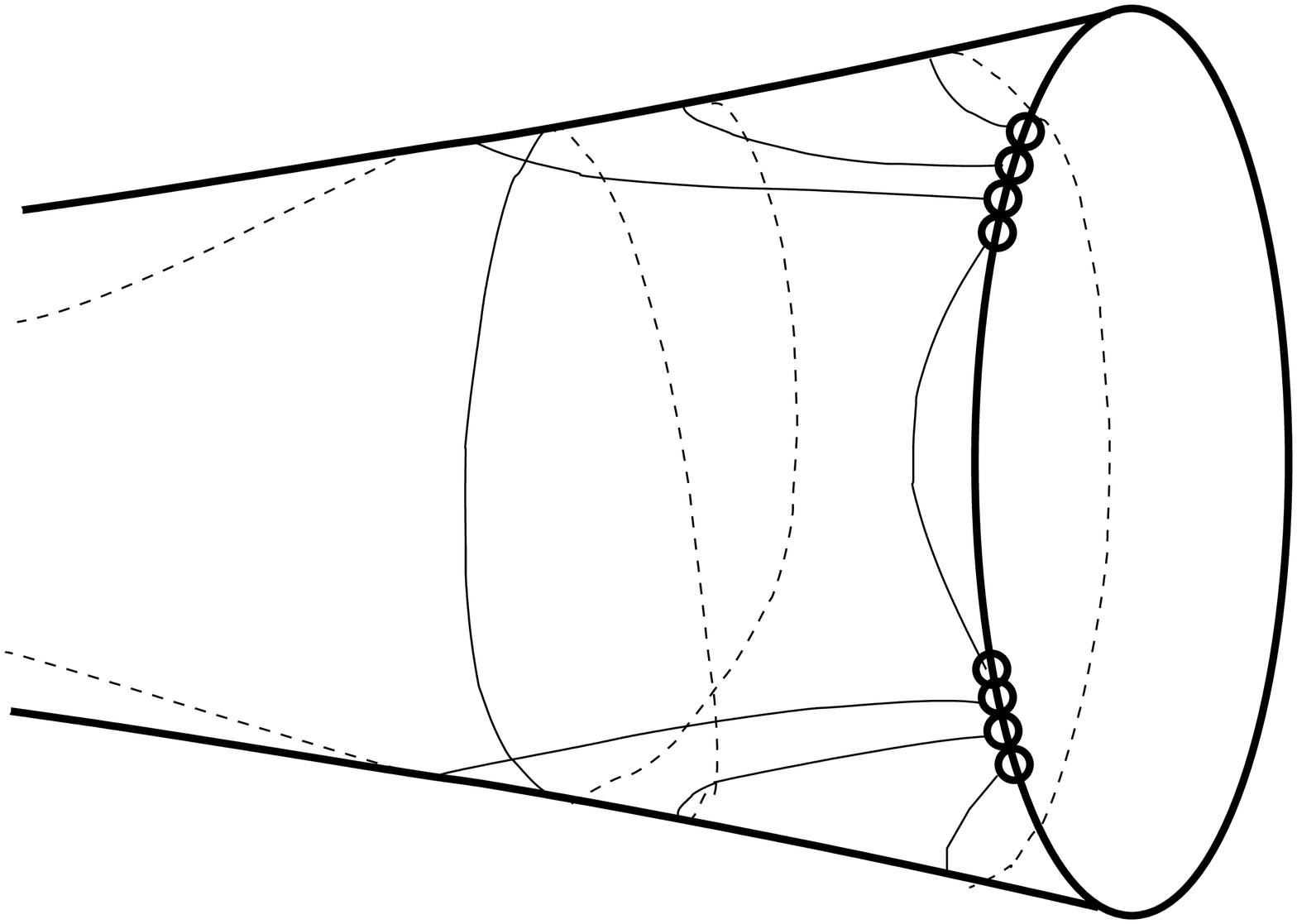}\\
\end{center}
\caption{Schematic figure of 
a contribution to 
$\langle 
\mbox{Tr}(\pa \Phi)^2 (\bpa \Phi)^2 
\mbox{Tr}(\pa \Phi)^2 (\bpa \Phi)^2 
\rangle_{S^1 \times R}$
coming from "string bits" 
propagating in the geometry I.
The $x$-direction is surpressed in the figure.
Small circles express "string bits" $\pa\Phi$ 
($\bpa \Phi$)
and lines
connecting the small circles are paths of the string bits.
When each bit is winding $\tau$ direction 
for different times, the string 
will be wildly torn apart
and closed string picture may not be adequate.}
\label{nc}
\end{figure}

\begin{figure}
\begin{center}
 \leavevmode
 \epsfxsize=60mm
 \epsfbox{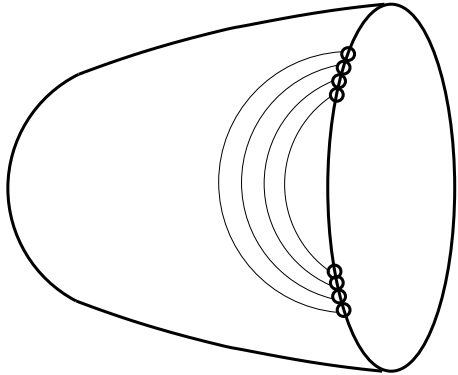}\\
\end{center}
\caption{The string bits propagating in the geometry II. 
In geometry II there is no non-contractible loop.}
\label{cc}
\end{figure}

\begin{figure}
\begin{center}
 \leavevmode
 \epsfxsize=60mm
 \epsfbox{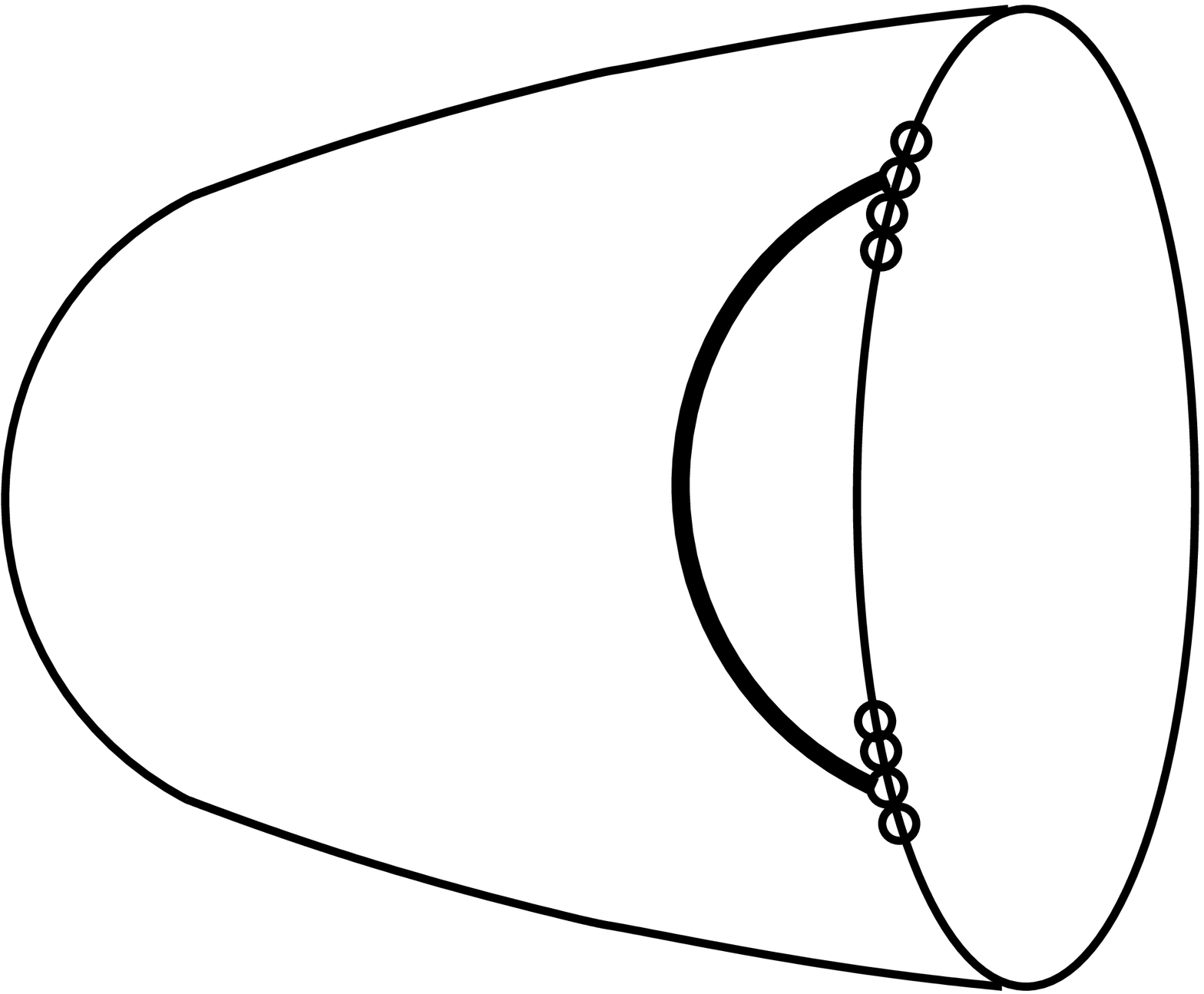}\\
\end{center}
\caption{One can "glue" string bits into a closed string
in the geometry II.}
\label{gl}
\end{figure}

Now let us examine 
the discussions of 
section \ref{Deconfined}
in this example.
We interprete
each field $\Phi$ as 
a string bit that makes up
a closed string.
Then the sum (\ref{g1z}) ((\ref{g1bz})) 
means in geometry I 
correlation functions are
obtained by summing 
over contributions of diagrams where
each bit winds the thermal 
circle for different times (Fig.\ref{nc}).
When each bit wind the thermal circle 
for different times,
string worldsheet interpretation 
may not be appropriate.
However, we can interpret
(\ref{g2z}) ((\ref{g2bz}))
as coming 
from a single propagator of bit
on a different geometry II. 
In geometry II 
there is no non-contractible circle
and hence there is no
summation over the winding modes (Fig.\ref{cc}).
Then those bits will be able to make up a closed string
worldsheet. 
We interprete multiplying field theory propagators
(\ref{g2z}) ((\ref{g2bz}))
as "gluing" of string bits into a closed string
(Fig.\ref{gl}).

The above picture is very much like the
interpretation of the Hagedorn transition
in string theory.
Indeed, when the 
spatial manifold
is some other manifold,
for example $S^3$,
the field theory
do have a Hagedorn density
of states and this phase
transition is at some 
non-zero temperature
\cite{Sundborg:1999ue,Polyakov:2001af,%
Aharony:2003sx,Aharony:2005bq}.\footnote{%
The large $N$ limit makes the phase
transition possible \cite{Gross:1980he,Wadia:1979vk}
on the compact space $S^3$.}
In this case the lightest mass of the fields
after Kaluza-Klein compactification
of the spatial manifold is most
relevant for determining
the Hagedorn temperature.
If we regard spatial line $R$ in our model
as an infinite interval limit of a segment
with Dirichlet boundary conditions
on both ends,
the Hagedorn temperature
is proportional to the
inverse of the length of the segment
and we may regard 
the zero-temperature phase transition
as a limit of the finite Hagedorn 
temperature.\footnote{%
In higher dimension, $S^1 \times R^{d-1}$
($d \geq 3$) can be regarded as
a limit of $S^1 \times S^{d-1}$ where
the radius of $S^{d-1}$ goes to infinity.
Then the zero-temperature
phase transition in $S^1 \times R^{d-1}$
can be regarded as a limit of the
Hagedorn transition in $S^1 \times S^{d-1}$.}

\section{Summary and Future Directions}

In this article we analyzed
the reorganization of field theory correlators
to closed string amplitudes
in Euclidean thermal field theory
and studied how the dual bulk geometry
is encoded on them.
The expectation value of
the Polyakov loop which is an
order parameter of confinement-deconfinement
phase is directly encoded on
the dual bulk geometry seen by
the field theory correlators.
Once the Polyakov expectation value is
correctly taken into account,
the gluing, which was a key step
for the reorganization of field theory
correlators to closed string amplitudes in 
\cite{Gopakumar:2004qb,Gopakumar:2005fx,Gopakumar:2004ys},
is straightforward in confined phase.
In deconfined phase the gluing was not
straightforward in momentum space.
We reexamined the meaning of the gluing
and argued that the gluing is still possible
if one correctly chooses the dual bulk geometry.
We presented free massless
scalar field theory on $S^1 \times R$
as a concrete realization of our arguments.
In our approach, the instability of a configuration
of the zero-mode of the
temporal gauge field $A_0$
at the point of phase transition
is directly {\em translated} 
into the instability
of the bulk geometry.

Studying other compactifications with
different field contents
is of course interesting.
Four dimensional 
Yang-Mills theory on $R^3$ 
is important 
for its direct relevance to the
real world.
Compactification to $S^3$
is also interesting.
Even at the weak coupling
the phase structure of 
it qualitatively 
resembles that of gravity, and
it may be continuously continued
to them in the strong coupling
\cite{Aharony:2003sx,Aharony:2005bq}
(see also \cite{Alvarez-Gaume:2005fv}). 
The dual geometry of confined phase
at low temperature
is thermal $AdS_5$, 
as we have discussed in general context.
Finding the geometry
which corresponds to deconfined phase
in this case at weak coupling
will not be so easy.
However, in the free field limit
it may still be possible to find
the geometry dual to the deconfined phase,
as suggested from the
tractability of the field theory side
in this limit.

Our analisis was in
the zero 't Hooft coupling limit,
and in this limit the string bits,
or $\Phi$ fields,
are just loosely 
tied together by Gauss' law constraint.
This corresponds to
a tensionless limit of
the dual closed string theory.
Turning finite 't Hooft coupling
will makes bits bind together
and their configuration will be more string like.
This corresponds to introducing a finite tension
in the dual closed string.
In flat space
the Hagedorn temperature
is governed by
string tension,
and it should be also 
relevant in the asymptotically
$AdS$ spaces.
In particular, 
the Hagedorn transition
and the phase transition in Euclidean gravity
(Hawking-Page transition \cite{Hawking:1982dh})
are observed to be separated
after including the finite coupling effect
\cite{Aharony:2003sx,Aharony:2005bq}.
It seems in all known cases at finite coupling
the Hawking-Page 
transition always occurs
before the Hagedorn transition,
but there is no general argument for
it must be so.\footnote{%
According to S. Minwalla in his lecture
at the Strings Meeting 2004 at Khajuraho.}
The indication of this fact to 
the similarity between  
the behavior of the string bits
and the interpretation of the
Hagedorn transition in string theory
is not clear to us yet.
But the identification of 
difficulty of the gluing with
the behavior of the string bits 
winding the thermal circle for
different times
applies no matter whether 
it is related to 
the Hagedorn transition or not.
The Hagedorn transition 
interpretation may still apply
through the interpretation of
the Hawking-Page 
transition as
a local Hagedorn transition,
as proposed in 
\cite{Barbon:2001di,Barbon:2002nw,Barbon:2004dd}.
It is very important to clarify those issues
at finite coupling.

It is also interesting to study
other modifications in 
the field theory side
and look for the
corresponding dual geometries.
For example,
massive deformation 
of field theory meets
obstruction
for the gluing 
procedure 
similar to the one we met 
in deconfined phase
if one works
in momentum space,
and one is urged to find
an appropriate deformation
in the dual geometry.
Studying massive case 
will be also useful
for applying our method
to the two-dimensional 
matrix models. 
In particular, 
it will be interesting to 
study the matrix model for
two dimensional black holes \cite{Kazakov:2000pm},
where the Polyakov loop also plays
key roles \cite{Suyama:2004vk}.
Also, recently in the context of
perturbative string 
theory in the orbifold of flat space
the authors of \cite{Costa:2005ej} identified
the mechanism of chronology protection
with Hagedorn-like transition
before closed null curves form.
On the other hand,
chronology protection had 
also been studied through 
holographic dual descriptions 
by boundary field theory (see the references in
\cite{Costa:2005ej}).
It will be interesting to relate
those two approaches 
through our method.

\vspace*{5mm}
\begin{center}
{\bf Acknowledgments}
\end{center}
Firstly, I must thank 
R. Gopakumar for
suggesting us 
the idea of probing
thermal geometry 
by his reorganization
of field theory correlators to closed strings,
as well as collaboration
in early stage of this work
and sharing his insights
through many discussions
throughout this work.
I am also grateful for 
his encouragements
and valuable comments on the manuscript.
I would also like to thank D. Astefanesei,
S. Bhattacharya and especially
K. P. Yogendran for collaboration
in early stage as well as stimulating 
discussions.
I am also thankful to 
J. R. David and
A. Sen for helpful discussions.
I am benefitted from 
the interactions in
Indian Strings Meeting 2004 at
Khajuraho, especially
I would like to thank
S. Minwalla for his insightful
lecture and discussions.
I would also like to thank 
the organizers, especially
A. Dhar, for the very nice workshop.
This research was made possible
by generous supports from people in India.

\vspace*{15mm}
\appendix
\noindent {\bf \Large Appendix}

\section{Geodesics in $AdS_{d+1}$}\label{A}

Here we give some formula for (regularized) 
geodesic distance
in $AdS_{d+1}$ used in section \ref{Example}.  
The method is fairly standard, see e.g.
\cite{Balasubramanian:1998de,Louko:2000tp}.

(Euclidean) $AdS_{d+1}$ can be described as
a surface in $d+2$ dimensional
flat space with signature $(-,+,\cdots,+)$:
\bea
\eta_{AB}y^Ay^B = - R^2 , \qquad
\eta_{AB} = \mbox{diag}(-,+,\cdots,+)  .
\eea
The geodesic distance $D(p,q)$
between points
$p$ and $q$ with coordinates
$y_p$ and $y_q$ respectively
is given by
\bea
D(p,q) = R \cosh^{-1} \frac{\langle y_p,y_q \rangle}{R^2}
\label{Dpq}
\eea
where $\langle y_p,y_q \rangle =
-\eta_{AB}y_p^Ay_q^B$.

Let us calculate the geodesic distance of
two points which are on the
boundary of the Poincare coordinates.
The coordinatization of Poincare coordinates
is given by
\bea
y^0 
&=& 
\frac{1}{2u} 
\left(1 + \frac{u^2}{2}(R^2+\sum_{i=1}^{d} (x^i)^2)
\right), \nonumber \\
y^{d}
&=& 
\frac{1}{2u} 
\left( -1 - \frac{u^2}{2}(R^2-\sum_{i=1}^{d} (x^i)^2)
\right), \nonumber \\
y^i &=&  R u x^i \qquad (i=1,\cdots, d)  .
\eea
Using (\ref{Dpq})
the geodesic distance between two points
$p = (u,x)$, $q = (u,0)$ at large $u$
turns out to be
\bea
D(p,q) \sim
R \log [\frac{R^2 u^4}{16} x^2].
\eea
Thus after subtracting
$x$ independent divergent piece,
the geodesic approximation gives
\bea
e^{-M D(p,q)} 
=
\left(
\frac{1}{x^2}
\right)^{MR}
\eea
where from the $AdS$/CFT dictionary 
the mass $M$ is given by
\cite{Gubser:1998bc,Witten:1998qj}
\bea
{M^2}{R^2} = \Delta(\Delta-d).
\eea
$\Delta$ is the dimension of 
corresponding operator.

Next let us study the static coordinates.
It is obtained by the following coordinate transformation
\bea
y^0 = \sqrt{r^2+R^2} \cosh \frac{t}{R}, \quad
y^1 = \sqrt{r^2+R^2} \sinh \frac{t}{R}, \quad
y^j = r \Omega^{j-2}_{d-1}
\eea
where $\Omega_{d-1}^j$ are angular coordinates on $S^{d-1}$,
we obtain the metric
\bea
ds^2
=
\left(
\frac{r^2}{R^2}+1
\right) dt^2
+
\left(
\frac{r^2}{R^2}+1
\right)^{-1}
dr^2
+ r^2 d\Omega_{d-1}^2  .
\eea
The $r \rightarrow \infty$ boundary is $S^3$.
The leading term of geodesic distance $D(p,q)$
between two points $p$ and $q$ in large $r$ is
\bea
{D(p,q)}
\sim
R \log \frac{r^2}{R^2}
\left(
\cosh \frac{t_1-t_2}{R} - \cos (\theta_1-\theta_2)
\right)  .
\eea
Thus by the saddle point (geodesic) approximation we get
\bea
e^{- M D(p,q)} \sim
\left(
\frac{r^2}{R^2}
\frac{1}{\cosh \frac{t_1-t_2}{R} - \cos (\theta_1-\theta_2)}
\right)^{MR}  .
\eea
By taking $r \rightarrow \infty$ limit
rescaling the divergent piece $\frac{r^2}{R^2}$,
we arrive at (\ref{cft2}) 
by identifying 
$\frac{t}{R}= \frac{x}{\beta}$, 
$\theta = \frac{\tau}{\beta}$,
in the large $J$ limit 
$MR = \sqrt{J(J-1)}\sim J + {\cal O}(1/J)$
for $d=2$.

\bibliography{thermal}
\bibliographystyle{kazu} 

\end{document}